\documentclass[10pt]{article}
\usepackage{nohyperref}
\usepackage{latexsym}
\usepackage{graphicx}
\usepackage{amsmath}
\usepackage{amssymb}
\usepackage{amsfonts}
\usepackage{epsfig}
\usepackage{fancybox}
\usepackage{natbib}
\usepackage{array}
\usepackage{subfig}
\usepackage{pstricks}
\usepackage{url}

\usepackage{algorithm}
\usepackage{algpseudocode}
\algdef{SxnE}[FOR]{For}{EndFor}[1]{\algorithmicfor\ #1 }

\graphicspath{{}}
\allowdisplaybreaks[3]

\newcommand{\head}[2]{\multicolumn{1}{>{\centering\arraybackslash}p{#1}}{\emph{#2}}}

\newcommand{\bcen}{\begin{center}}
\newcommand{\ecen}{\end{center}}

\newcommand{\beqn}{\begin{equation}}
\newcommand{\eeqn}{\end{equation}}
\newcommand{\beqns}{\begin{equation*}}
\newcommand{\eeqns}{\end{equation*}}

\newcommand{\beqnary}{\begin{eqnarray}}
\newcommand{\eeqnary}{\end{eqnarray}}
\newcommand{\beqnarys}{\begin{eqnarray*}}
\newcommand{\eeqnarys}{\end{eqnarray*}}

\newcommand{\bary}{\begin{array}}
\newcommand{\eary}{\end{array}}

\newcommand{\benm}{\begin{enumerate}}
\newcommand{\eenm}{\end{enumerate}}

\newcommand{\bitem}{\begin{itemize}}
\newcommand{\eitem}{\end{itemize}}

\newcommand{\mbR}{\mathbb{R}}

\def\cA{{\cal A}}

\def\cS{{\cal S}}
\def\cT{{\cal T}}

\def\cV{{\cal V}}

\newcommand{\bI}{{\bf I}}

\renewcommand{\r}{{\bf r}}

\newcommand{\bX}{{\bf X}}
\newcommand{\bx}{{\bf x}}

\newcommand{\by}{{\bf y}}

\newcommand{\beps}{\mbox{\boldmath{$\varepsilon$}\unboldmath}}
\newcommand{\bmu}{\mbox{\boldmath{$\mu$}\unboldmath}}

\newcommand{\bzero}{\mathbf{0}}
\newcommand{\bbeta}{\mbox{\boldmath{$\beta$}}}
\newcommand{\bEta}{\mbox{\boldmath{$\eta$}}}

\newcommand{\lam}{\lambda}

\newcommand{\ra}{\rightarrow}

\newcommand{\overt}{\frac{1}{2}}

\newcommand{\overn}{\frac{1}{n}}

\newcommand{\veps}{\varepsilon}

\begin{document}

\begin{center}
{\Large \bf Strong rules for nonconvex penalties and their implications for efficient algorithms in high-dimensional regression}\\[10pt]
\vspace{0.3cm} \textrm{Sangin Lee and} \textrm{Patrick Breheny}
\\[0pt]
\vspace{0.1cm} \textit{The University of Iowa}\\[0pt]
\end{center}

\begin{abstract}
We consider approaches for improving the efficiency of algorithms for
fitting nonconvex penalized regression models such as SCAD and MCP in
high dimensions.  In particular, we develop rules for discarding
variables during cyclic coordinate descent.  This dimension reduction
leads to a substantial improvement in the speed of these algorithms
for high-dimensional problems.  The rules we propose here eliminate a
substantial fraction of the variables from the coordinate descent
algorithm.  Violations are quite rare, especially in the locally
convex region of the solution path, and furthermore, may be easily
detected and corrected by checking the Karush-Kuhn-Tucker conditions.
We extend these rules to generalized linear models, as well as to
other nonconvex penalties such as the $\ell_2$-stabilized Mnet
penalty, group MCP, and group SCAD.  We explore three variants of the
coordinate decent algorithm that incorporate these rules and study the
efficiency of these algorithms in fitting models to both simulated
data and on real data from a genome-wide association study.
\end{abstract}

\medskip
\noindent {\em Keywords}: Coordinate descent algorithms, Local convexity,
Nonconvex penalties, Dimension reduction.

\section{Introduction}\label{introduction}

Consider the linear regression model
\begin{equation}\label{linear:model}
    \by=\bX\bbeta+\beps,
\end{equation}
where $\by=(y_1,\cdots,y_n)'$ is the vector of $n$ response variables,
$\bX=(\bx_1,\dots,\bx_p)$ is the $n\times p$ design matrix with the
$j$th column $\bx_j=(x_{1j},\dots,x_{nj})'$, $\bbeta=(\beta_1, \dots,
\beta_p)'$ is the vector of regression coefficients and
$\beps=(\veps_1,\cdots,\veps_n)'$ is the vector of random errors. We
assume that the responses and covariates are centered so that the
intercept term is zero. We are interested in estimating the vector of
regression coefficients $\bbeta$. Penalized regression methods
accomplish this by minimizing an objective function $Q$ that is
composed of the sum of squared residuals plus a penalty. The penalized
least squares estimator is defined as the minimizer of
\begin{equation}\label{obj:non}
    Q_{\lambda,\gamma}(\bbeta)=\frac{1}{2n}\|\by-\bX\bbeta\|_2^2 + \sum_{j=1}^{p}
    J_{\lambda,\gamma}(|\beta_j|),
\end{equation}
where $J_{\lambda,\gamma}(\cdot)$ is a penalty function indexed by a
regularization parameter $\lambda$ that controls the balance between the
fit of the model and the penalty, and the penalty function may depend
on one or more tuning parameters $\gamma$.

Here we focus on optimization algorithms for penalized regression
methods. There has been much work on developing efficient algorithms
for many problems with various penalties, including
\cite{Efron2004a}, \cite{friedman2007pathwise}, and \citet{Wu2008}
for the least absolute selection operator (LASSO), and
\cite{kim2008smoothly}, \cite{zou2008one}, and
\cite{breheny2011coordinate} for nonconvex penalties such as
smoothly clipped absolute deviation (SCAD) and the minimax concave
penalty (MCP). Recently, several authors have investigated rules for
discarding variables during certain steps of the above algorithms,
thereby saving computational time through dimension reduction.

For the LASSO, \cite{el2011safe} proposed the basic SAFE rule discards
the $j$th variable if
\begin{equation} \label{safe:rule}
    |\bx_j'\by/n| < \lambda - \frac{1}{n}\|\bx_j\|_2\|\by\|_2 (\lambda_{\max}-\lambda)/\lambda_{\max},
\end{equation}
where $\lambda_{\max}=\max_j |\bx_j'\by/n|$ is the smallest tuning
parameter value for which all estimated coefficients are zero. They
proved that the estimated coefficient for any variable satisfying the
basic SAFE rule (\ref{safe:rule}) must be zero in the solution at
$\lambda$. \cite{tibshirani2012strong} proposed the basic strong rule
by modifying the basic SAFE rule (\ref{safe:rule}). For a standardized
design matrix ($\|\bx_j\|_2/\sqrt{n}=1$ for all $j$), we have
$\|\by\|_2/\sqrt{n} > \lambda_{\max}$ by the Cauchy-Schwarz inequality
and therefore $2\lambda - \lambda_{\max}$ is an upper bound of the
quantity on the right hand side of \eqref{safe:rule}.  The strong rule
therefore discards the $j$th variable if
\begin{equation} \label{strong:rule}
    |\bx_j'\by/n| < 2\lambda - \lambda_{\max}.
\end{equation}
Being an upper bound of the SAFE rule, the strong rule
(\ref{strong:rule}) discards more variables than the SAFE rule.
Unlike the SAFE rule, however, it is possible for the strong rule to
be violated.  Because strong rules can mistakenly discard active
variables (i.e., variables whose solution is nonzero for that value of
$\lambda$), \cite{tibshirani2012strong} proposed checking the
discarded variables against the Karush-Kuhn-Tucker (KKT) conditions to
correct for any violations that may have occurred during the
optimization.

These basic rules are most useful at large values of $\lambda$ and
rarely eliminate variables at smaller $\lambda$ values.  This is
unfortunate from an algorithmic perspective, since the majority of
time required to fit a regularization path is spent during
optimization for the small $\lambda$ values.  To overcome this
drawback, \cite{tibshirani2012strong} proposed sequential strong
rules. For a decreasing sequence of tuning parameter $\lambda_1 \geq
\lambda_2 \geq \dots \geq \lambda_m$, the sequential strong rule
discards the $j$th variable from the optimization problem at
$\lambda_k$ if
\begin{equation} \label{seq:strong:rule}
    |\bx_j'\r_{k-1}/n| < 2\lambda_k -
    \lambda_{k-1},
\end{equation}
where $\r_{k-1} = \by-\bX\hat\bbeta(\lambda_{k-1})$ is the vector of
residuals at $\lam_{k-1}$. Unlike the basic rules, the sequential
strong rule discards a large proportion of inactive variables at all
values of $\lambda$.  In addition, the rule is rarely violated, and
is therefore unlikely to discard active variables by mistake.

In this paper, we investigate sequential strong rules for discarding
variables in penalized regression with nonconvex penalties, as well as
strategies for incorporating these rules into coordinate descent
algorithms for fitting these models.  In addition, we derive rules for
discarding variables in various related problems with nonconvex
penalties, such generalized linear models, $\ell_2$-stabilized
penalties (the ``Mnet'' estimator), and grouped penalties.  We provide
a publicly available implementation of these algorithms in the updated
\verb"ncvreg" package (available at \url{http://cran.r-project.org}),
which was used to fit all the models in this paper.


\section{Strong rules in nonconvex penalized regression}\label{strong:non}

The basic idea of the sequential strong rule
\citep{tibshirani2012strong} is that the solution path
$\hat{\bbeta}(\lam)$ is a continuous function; furthermore, one can
obtain an approximate bound on how fast the solution path can change
as a function of lambda.  Thus, when solving for $\hat{\bbeta}$ at
$\lam_{k-1}$ and then again at $\lam_k$, we can exclude certain
variables from the optimization procedure because they aren't close
enough to the threshold for inclusion in the model to reach that
threshold in the distance between $\lam_{k-1}$ and $\lam_k$.  The
effect is that the dimension of the optimization problem is reduced --
instead of cycling over all variables, the estimation procedure needs
only to cycle over a much smaller set of variables capable of entering
the model at $\lam_k$.

The bound investigated by \citep{tibshirani2012strong} is given by
\begin{equation}\label{unit:slope}
    |c_j(\lambda)-c_j(\tilde\lambda)| \leq |\lambda-\tilde\lambda|,
    \text{ for any } \lambda \text{ and } \tilde\lambda,
\end{equation}
where $c_j(\lambda)=\bx_j'\r(\lam)/n$ is the
correlation\footnote{Strictly speaking, $c_j(\lam)$ is not a
  correlation, since $\r$ is not standardized, and thus only
  proportional to the correlation between $\bx_j$ and $\r$.  However,
  the term is widely used; see, e.g., \citet{Efron2004a}.} between
variable $j$ and the residual at $\lam$.  This condition is equivalent
to $c_j(\lambda)$ being continuous everywhere, differentiable almost
everywhere, and satisfying $|\nabla c_j(\lambda)|\leq 1$ wherever this
derivative exists.  Tibshirani et al.~called condition
(\ref{unit:slope}) the unit slope bound.  If condition
(\ref{unit:slope}) holds, then for any variable $j$ satisfying the
sequential strong rule (\ref{seq:strong:rule}), we have
$c_j(\lambda_k)<\lambda_k$, and thus, $\hat\beta_j(\lambda_k)=0$ by
the KKT conditions of the LASSO.


In this section, we examine whether a variation of condition
\eqref{unit:slope} holds for the MCP and SCAD penalties, and use a
modified version of \eqref{unit:slope} to develop strong rules for
those nonconvex penalties.  Also, we provide numerical examples to
illustrate the application of the strong rules on a simulated data
set.  Lastly, we note that unlike the LASSO case, for nonconvex
penalties the function $c_j(\lam)$ is not guaranteed to be continuous;
we explore the consequences of this fact in
Section~\ref{Sec:local-convexity}.

\subsection{MCP}
\label{Sec:strong-mcp}
\cite{zhang2010nearly} proposed the MCP which is defined as
\begin{align*}
    J_{\lambda,\gamma}(t)=
    \begin{cases}
        -t^2/(2\gamma)+\lambda t, & \text{if } t\leq\gamma\lambda, \\
        \gamma\lambda^2/2, & \text{if } t>\gamma\lambda.
    \end{cases}
\end{align*}
for $\lambda\geq0$ and $\gamma>1$. We begin by noting the KKT
conditions for the penalized problem (\ref{obj:non}),
\begin{align}
\begin{aligned}\label{kkt:con}
  \bx_j'&\r/n = \nabla J_{\lambda}(|\hat\beta_j|) &&\text{ for all $j \in \cA$},\\
  |\bx_j'&\r/n| < \lambda &&\text{ for all $j \notin \cA$}
\end{aligned}
\end{align}
where $\cA=\{j:\hat\beta_j \neq 0\}$ is the active set.

Variables in $\cA$ are continuously changing as a function of $\lam$,
but many variables in $\cA^c$ remain zero from one $\lam$ value to the
next.  Our aim, then, is to develop a screening rule to can discard
the variables in the inactive set $\cA^c$ that are likely to remain
zero.  In high dimensions, doing so should yield substantial
computational savings. From the KKT conditions (\ref{kkt:con}), we
have the form of $c_j(\lambda)$,
\begin{align}\label{grd:mcp}
    c_j(\lambda)=
    \begin{cases}
        0, & \text{if } |\hat\beta_j|>\gamma\lambda, \\
        -\hat\beta_j/\gamma + \lambda \text{sign}(\hat\beta_j) & \text{if } |\hat\beta_j|\leq\gamma\lambda,~\hat\beta_j\neq0 \\
        \bx_j'\bX_\cA (\bX'_\cA\bX_\cA)^{-1}\nabla J_{\lambda}(|\hat\bbeta_\cA|) + (Const), & \text{if } \hat\beta_j=0,
    \end{cases}
\end{align}
where $\bbeta_\cA=(\beta_j, j\in\cA)$ and $\bX_\cA=(\bx_j, j\in\cA)$
denote the subvector and submatrix of $\bbeta$ and $\bX$,
respectively, and $(Const)$ stands for constant terms not depending on
$\lambda$.  Unlike the LASSO, the above expression for $c_j(\lambda)$
does not permit a closed-form expression for $\nabla c_j(\lambda)$ for
variables in the active set.  Hence, we investigate an approximation
for $\nabla c_j(\lambda)$ based on an orthogonal design matrix.  In
this case, the coefficient estimates have closed form solution
$\hat\beta_j = \frac{\gamma}{\gamma-1} \text{sign}(z_j)
(|z_j|-\lambda)_{+}$, where $z_j=\bx'_j\by/n$ is the ordinary least
squares estimator, and the second term of (\ref{grd:mcp}) is
$c_j(\lambda) = \text{sign}(z_j) \left\{\lambda - \frac{1}{\gamma-1}
(|z_j|-\lambda)_{+}\right\}$.  This suggests the bound $|\nabla
c_j(\lambda)| \leq 1+1/(\gamma-1)$.  This slope bound is larger than
the corresponding bound for the LASSO, as the nonconvexity of MCP
allows the solution path -- and thus, $c_j(\lam)$ -- to change more
rapidly as a function of $\lam$ than it does for LASSO.  Note that in
the limiting case $\gamma\ra\infty$, MCP is equal to the lasso
penalty, and the bounds coincide.  Conversely, as $\gamma\ra 1$, MCP
is equivalent to hard thresholding.  The bound diverges in this case,
and there is no limit to the rate at which the solution path may
change and no possibility of discarding variables based on this
argument.

As in the LASSO case, a slope bound for variables in the active set
does not necessarily extend to variables in the inactive set.
Nevertheless, it is reasonable to expect that the correlation with the
residuals is changing more rapidly for variables in the active set
than variables in the inactive set.  This line of thinking that allows us
to establish an explicit rule for screening predictors during
optimization.

If, for $j=1,\dots,p$, the bound
\begin{equation}\label{mcp:slope}
    |c_j(\lambda)-c_j(\tilde\lambda)| \leq \frac{\gamma}{\gamma-1}|\lambda-\tilde\lambda|,
    \text{ for any } \lambda \text{ and } \tilde\lambda,
\end{equation}
holds, we can obtain the following rule, which we call the (sequential) strong rule for MCP:
\begin{equation}\label{rule:mcp}
    |\bx_j'\r_{k-1}/n| < \lambda_k + \frac{\gamma}{\gamma-1}(\lambda_k
    - \lambda_{k-1}).
\end{equation}
Note that, for any variable $j$ satisfying \eqref{mcp:slope} and
\eqref{rule:mcp}, we have
\begin{eqnarray*}
    |c_j(\lambda_k)| &\leq& |c_j(\lambda_k)-c_j(\lambda_{k-1})|+|c_j(\lambda_{k-1})| \\
    &<& \frac{\gamma}{\gamma-1}(\lambda_{k-1}-\lambda_k)+\lambda_k+\frac{\gamma}{\gamma-1}(\lambda_k - \lambda_{k-1})\\
    &=& \lambda_k,
\end{eqnarray*}
and thus, $\hat\beta_j(\lambda_k)=0$.

Indeed, as we shall see, the heuristic argument that residual
correlation changes more rapidly in the active set than the inactive
set holds up quite well in practice.  Nevertheless, violations are
possible, and thus it is necessary to check the discarded variables
against the KKT conditions \eqref{kkt:con} as a final step in the
optimization algorithm.

\subsection{SCAD}
The SCAD penalty proposed by \cite{fan2001variable} is defined as
\begin{align*}
    J_{\lambda,\gamma}(t)=
    \begin{cases}
        \lambda t, & \text{if } t\leq\lambda, \\
        \{\gamma\lambda(t-\lambda)-(t^2-\lambda^2)/2\}/(\gamma-1), & \text{if } t\leq\gamma\lambda, \\
        (\gamma-1)\lambda^2/2+\lambda^2, & \text{if } t>\gamma\lambda.
    \end{cases}
\end{align*}
for $\lambda\geq0$ and $\gamma>2$. From the KKT conditions
(\ref{kkt:con}), we have
\begin{align*}
    c_j(\lambda) =
    \begin{cases}
        0, & \text{if } |\hat\beta_j|>\gamma\lambda \\
        (\gamma\lambda\text{sign}(\hat\beta_j)
        - \hat\beta_j)/(\gamma-1), & \text{if }\lambda<|\hat\beta_j|\leq\gamma\lambda \\
        \lambda \text{sign}(\hat\beta_j), & \text{if } |\hat\beta_j|\leq\lambda,~\hat\beta_j\neq0 \\
        \bx_j'\bX_\cA (\bX^T_\cA\bX_\cA)^{-1}\nabla J_{\lambda}(|\hat\bbeta_\cA|) + (Const), & \text{if }\hat\beta_j=0,
    \end{cases}
\end{align*}
where $(Const)$ stands for constant terms not depending on
$\lambda$. For SCAD, the orthogonal design solution is $\hat\beta_j =
\text{sign}(z_j)(\frac{\gamma-1}{\gamma-2}) \{|z_j|-\lambda\gamma /
(\gamma-1)\}_{+}$.  Applying the same reasoning as in
Section~\ref{Sec:strong-mcp}, we obtain the approximate slope bound
$|\nabla c_j(\lambda)| \leq 1+2/(\gamma-2)$ and the sequential strong rule
for SCAD
\begin{equation}\label{rule:scad}
    |\bx_j'\r_{k-1}/n| < \lambda_k +
    \frac{\gamma}{\gamma-2}(\lambda_k-\lambda_{k-1}).
\end{equation}
Like MCP, the SCAD solution path is capable of changing more rapidly
with respect to $\lam$ than LASSO, and thus requires a larger bound
for its strong rule.

\subsection{Numerical illustrations}

We now provide an illustration of the how the strong rules perform
using a simulated example.  The design of the simulation, which we
also use for the simulation study in Section~\ref{simulation}, is as
follows. All covariates marginally follow standard Gaussian
distributions, with a common correlation $\rho$ between any two
covariates. The response variable $y$ is generated from the linear
model (\ref{linear:model}) with errors drawn from the standard
Gaussian distribution. For each independently generated data set, we
set $n=200$ and $p=2,000$, with $20$ nonzero coefficients set to be
$\pm 1$ for linear regression and the remaining $1,980$
coefficients equal to zero.  Throughout this paper, we fix $\gamma=3$
for MCP and $\gamma=4$ for SCAD, roughly in line with recommendations
suggested in \cite{fan2001variable} and \cite{zhang2010nearly},
respectively.

\begin{figure}[ht!]
 \centering
 \includegraphics[width=\linewidth]{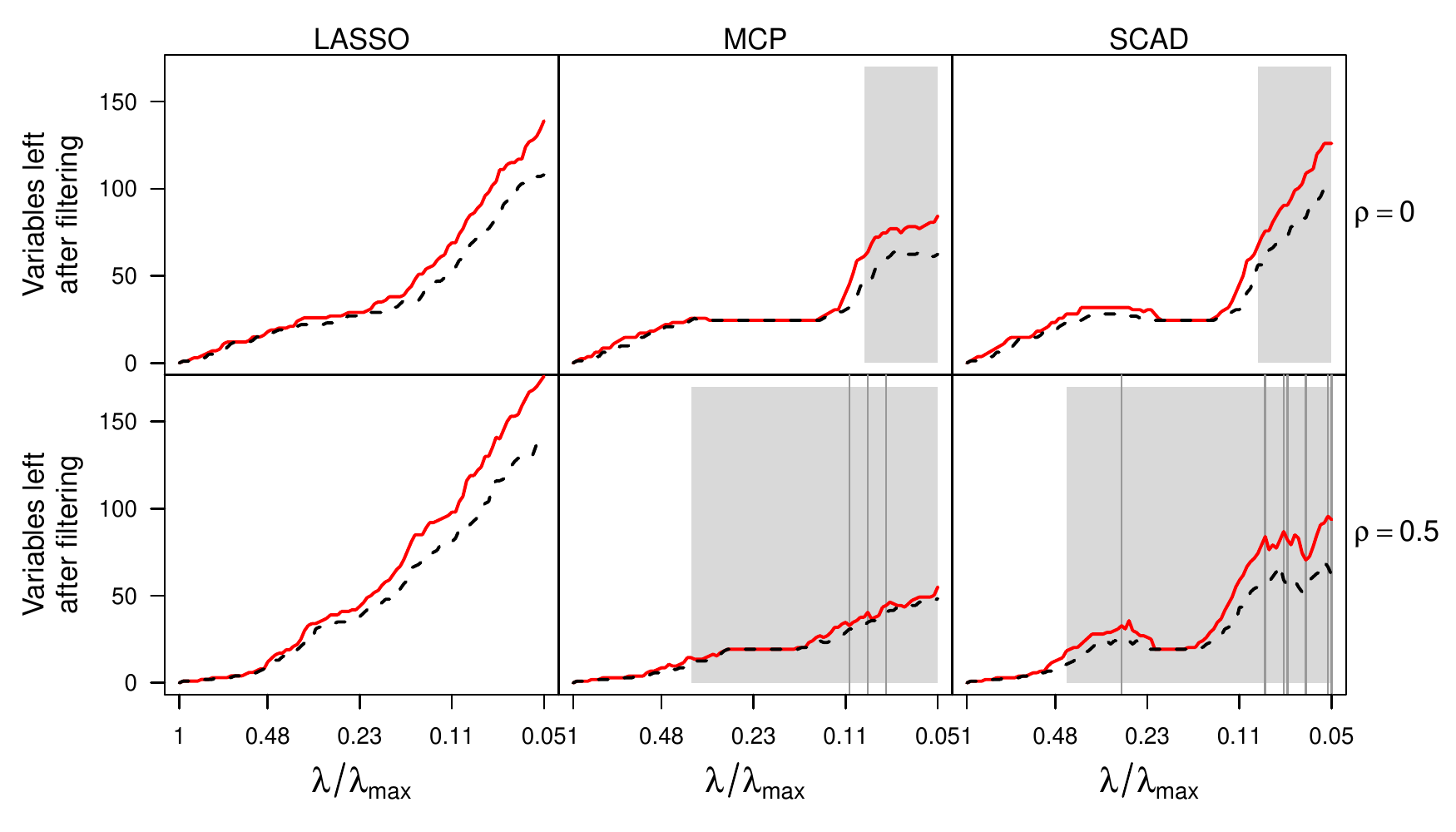}
 \caption{\label{loc-con} Application of the strong rules
   (\ref{seq:strong:rule}, \ref{rule:mcp}, and \ref{rule:scad}) for
   two simulated data sets, one with $\rho=0$ (top panel) and the
   other with $\rho=0.5$ (bottom panel).  The solid line is the number
   of variables left after filtering by strong rules and the dotted
   line is the actual number of active variables for each $\lambda$.
   The region of the coefficient path that does not satisfy local
   convexity is shaded gray.  Vertical lines are drawn for any value
   of $\lam$ at which a violation of the strong rules occurred.}
\end{figure}

Figure~\ref{loc-con} displays the performance of the strong rules for
LASSO, MCP, and SCAD on two simulated data sets, one with uncorrelated
covariates, the other with a pairwise correlation of $\rho=0.5$.  The
figure displays the number of variables remaining (i.e., $p$ minus the
number of discarded variables) after applying the strong rules for a
decreasing sequence of $\lam$ values, alongside the actual number of
nonzero coefficients in the model for those $\lam$ values.  Vertical
lines are drawn for each value of $\lam$ for which a violation of the
strong rules occurred.  So in this example, there were no violations
for any of the methods when $\rho=0$, but we observe 3 violations for
MCP and 7 violations for SCAD when $\rho=0.5$.

The strong rules perform remarkably well here, especially for
$\rho=0$.  The vast majority of the $p=2,000$ variables are discarded
by the strong rules.  In fact, nearly all of the variables that should
be discarded are discarded, across the entire path of $\lam$ values.
With so many variables discarded by the strong rules, it is surprising
how rare it is for a variable to be erroneously discarded.

Nevertheless, violations do occur, and are more common for nonconvex
penalties than for the LASSO, as we discuss in the next section.
Violations are important, but not fatal -- an algorithm based on
dimension reduction through discarding variables can always check the
validity of the dimension reduction by inspecting the KKT conditions
for $\hat{\bbeta}(\lam)$ upon convergence for each value of $\lam$,
and include any variables that were erroneously discarded.  In this
manner, we ensure that all solutions $\hat{\bbeta}$ returned by the
algorithm are indeed a (local) minimum of the objective function.
Details for constructing algorithms based on strong rules are given in
Section~\ref{algorithm}.

Although none occurred in this example, violations are also possible
for the lasso, and a similar KKT-checking step is required in the
LASSO algorithm proposed by \citet{tibshirani2012strong}.  A
systematic numerical study of the frequency of violations for MCP and
SCAD is provided in Section~\ref{simulation}.

\subsection{Local convexity}
\label{Sec:local-convexity}
Unlike the LASSO solution path, for nonconvex penalties $\hat{\bbeta}$
is not necessarily a continuous function of $\lam$.  It is possible
for the objective function to possess multiple local minima, and for
$\hat{\bbeta}(\lam)$ to ``jump'' from one local minimum to a different
local minimum between $\lam_{k-1}$ and $\lam_k$.  Such a discontinuity
undermines the entire premise of strong rules.



It is possible, however, to characterize the regions of the solution
path where such discontinuities may and may not occur.  The portion of
the solution path guaranteed to be continuous was referred to in
\citet{breheny2011coordinate} as the locally convex region.  Letting
$\cA(\lambda)=\{j:\hat\beta_j(\lambda)\neq0\}$ denote the active set
of variables at $\lambda$, and $\tau_{\min}(\lambda)$ denote the
minimum eigenvalue of $\bX_{\cA(\lambda)}'\bX_{\cA(\lambda)}/n$, the
solution path is said to be locally convex at $\lam$ if $\gamma >
1/\tau_{\min}(\lambda)$ for MCP, and $\gamma > 1 + 1 /
\tau_{\min}(\lambda)$ for SCAD.  Correspondingly, the locally convex
region is defined as $(\lam_{\max}, \lam^*)$, where $\lam^*$ is the
first (i.e., largest) value of $\lam$ for which the solution is no
longer locally convex.  As demonstrated in
\citet{breheny2011coordinate}, coefficient paths for nonconvex
penalties are smooth and well behaved in the locally convex region,
but may be discontinuous and erratic in the non-locally convex region.



We would therefore expect strong rules to be less reliable in the
non-locally convex region, and this is precisely what we see in
Figure~\ref{loc-con}, where all the observed violations occur in the
non-locally convex region.  Indeed, although this is not apparent in
the figure, several violations tend to occur simultaneously when a
discontinuity arises in the solution path.  For example, at
$\lam=0.0745$, there were 11 variables excluded by the strong rules
that were discovered during the KKT check to be nonzero at the new
local minimum $\hat{\bbeta}(\lam)$.

The presence of discontinuities in the solution paths for nonconvex
penalties places an inherent limitation on the use of sequential rules
to improve optimization efficiency during model fitting.
Nevertheless, as we will see, even in highly correlated settings, only
a small number of $\lam$ values experience violations, and strong
rules may be profitably incorporated into optimization algorithms for
nonconvex penalized models despite these violations, solving for the
solution path $\hat{\bbeta}(\lam)$ substantially faster than cyclic
coordinate descent approaches.


\section{Extensions to other nonconvex penalized models}\label{sec-var}

\subsection{$\ell_2$-stabilization}

To stabilize the solution path for nonconvex penalties, especially in
$p>n$ problems with highly correlated predictors,
\cite{huang2013balancing} proposed the Mnet estimator, which is
defined as the minimizer of
\begin{equation}\label{def:mnet}
    Q_{\lambda,\gamma}(\bbeta)=\frac{1}{2n}\|\by-\bX\bbeta\|_2^2 + \sum_{j=1}^{p}J_{\lambda_1,\gamma}(|\beta_j|)
    + \overt\lambda_2\sum_{j=1}^{p}\beta_j^2,
\end{equation}
where $J_{\lambda_1,\gamma}(\cdot)$ is the MCP. The logic behind the
estimator is the same as that of the elastic net
\citep[or Enet, ][]{Zou2005}, but with MCP replacing the LASSO in the
penalty.  Let
\begin{equation*}
    \tilde\bX=\left(\begin{array}{c} \bX \\ \sqrt{n\lambda_2}\,
    \bI_p \end{array} \right),~ \tilde\by=\left(\begin{array}{c} \by \\ \bzero_p \end{array} \right),
\end{equation*}
where $\bI_p$ is the $p \times p$ identity matrix and $\bzero_p$ is
the $p$-dimensional vector whose all elements are zero. Then the
criterion (\ref{def:mnet}) may be rewritten as
\begin{equation}
    \frac{1}{2n}\|\tilde\by-\tilde\bX\bbeta\|_2^2 + \sum_{j=1}^{p}J_{\lambda_1,\gamma}(|\beta_j|).
\end{equation}
Hence, we can directly apply the sequential strong rule
(\ref{rule:mcp}) to discard variables. Reparameterizing the problem in
terms of $\lambda_1=\alpha\lambda$ and $\lambda_2=(1-\alpha)\lambda$,
the strong rule for Mnet becomes
\begin{equation*}\label{rule0:mnet}
    \left|-\bx_j'\r_{k-1}/n+\lambda_k(1-\alpha)\hat\beta_j(\lambda_{k-1})\right|
    < \alpha\left\{\lambda_k + \frac{\gamma}{\gamma-1}(\lambda_k-\lambda_{k-1})\right\},
\end{equation*}
since $\tilde\bx_j'\tilde\by=\bx_j'\by$ and
$\tilde\bx_j'\tilde\bX\hat\bbeta = \bx_j'\bX\hat\bbeta +
n\lambda(1-\alpha)\hat\beta_j$. For inactive variables
($\hat{\beta}_j=0$), the above rule reduces to
\begin{equation}\label{rule1:mnet}
    \left|\bx_j'\r_{k-1}/n\right| < \alpha\left\{\lambda_k +
    \frac{\gamma}{\gamma-1}(\lambda_k-\lambda_{k-1})\right\}.
\end{equation}

\begin{figure}[ht!]
 \centering
 \includegraphics[width=\linewidth]{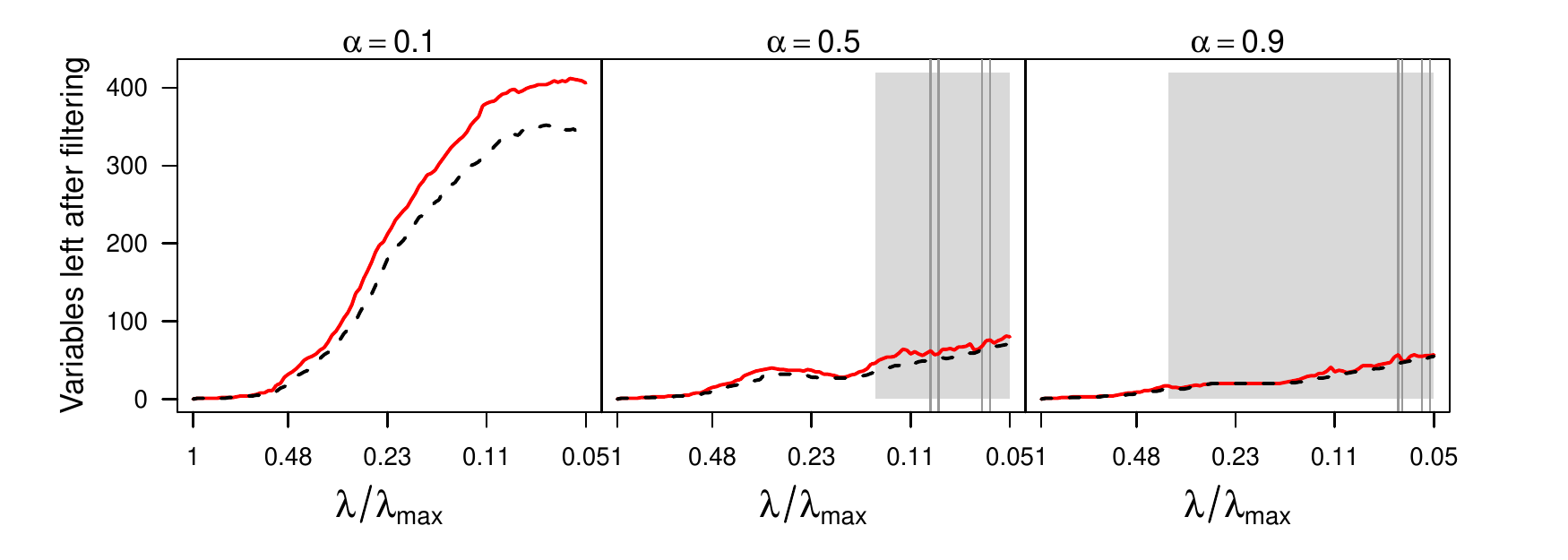}
 \caption{\label{mnet} Application of strong rule \eqref{rule1:mnet}
   to a simulated data set with $\rho=0.5$ for different values of
   $\alpha$.  As in Figure~\ref{loc-con}, the solid line is the number
   of variables left after filtering by strong rules, the dotted line
   is the actual number of active variables for each $\lambda$, the
   region of the coefficient path that does not satisfy local
   convexity is shaded gray, and vertical lines are drawn for any
   value of $\lam$ for which a violation of the strong rule occurred.}
\end{figure}

Figure~\ref{mnet} illustrates the application of strong rules to the
Mnet estimator for a simulated data set with $\rho=0.5$ as we vary the
parameter $\alpha$ that controls the MCP/$\ell_2$ balance in the
penalty.  As in Section~\ref{Sec:local-convexity}, one may
characterize the locally convex region; for the Mnet estimator, this
consists of the values of $\lambda$ satisfying
$\gamma>1/\{\tau_{\min}(\lambda) + (1-\alpha)\lambda\}$.

From Figure~\ref{mnet}, we can see that as we decrease $\alpha$ and
thereby increase the $\ell_2$ proportion of the penalty, the locally
convex region is extended, the model becomes less sparse, and fewer
issues with discontinuities and strong rule violations arise.  Indeed,
at $\alpha=0.1$, the objective function is locally convex over the
entire solution path and no violations occurred.  For all $\alpha$
values, the strong rules were successful in discarding a large
proportion of the inactive variables.

\subsection{Generalized linear models}\label{glm}
Suppose that the distribution of $\by|\bX$ falls within the framework
of the generalized linear model (GLM), with link function
$\eta_i=g(\mu_i)$, where $\mu_i=E(y_i|x_{i1}, \ldots, x_{ip})$ and 
\begin{equation}\label{glm:lp}
    \bEta=\bX\bbeta=\beta_0+\bx_1\beta_1+\dots +\bx_p\beta_p,
\end{equation}
where $\bEta=(\eta_1,\ldots,\eta_n)'\in\mbR^n$ is the vector of linear
predictors.
The nonconvex penalized estimator for a GLM is defined as the
minimizer of the negative log-likelihood plus the penalty term.  For
example, for logistic regression,
\begin{equation}\label{log:obj}
    -\overn\sum_{i=1}^{n}\left\{y_i\log \mu_i + (1-y_i)\log(1-\mu_i)\right\} + \sum_{j=1}^{p} J_{\lambda,\gamma}(|\beta_j|).
\end{equation}

The extension of strong rules to GLMs is straightforward. Indeed, both
the KKT conditions \eqref{kkt:con} and the strong rules themselves
(\ref{rule:mcp}, \ref{rule:scad}) are the same as in the linear case,
although the residual vector must now incorporate the link function:
$\r = \by - \bmu(\bEta)$.  For example,
$\bmu(\bEta)=\exp(\bEta)/(1+\exp(\bEta))$ for logistic regression and
$\bmu(\bEta)=\exp(\bEta)$ for Poisson regression.




\subsection{Nonconvex group penalized estimation}\label{sec:group}
Nonconvex penalties have also been proposed in the context of group
variable selection.  Suppose that the covariates may be grouped into
$G$ groups, with the grouping structure non-overlapping and known in
advance:
\begin{equation}\label{eqn:group}
    \by=\sum_{g=1}^G \bX_g\bbeta_g + \beps.
\end{equation}
where $\bX_g$ is the $n\times p_g$ design matrix corresponding to the
$g$th group and $\bbeta_g \in \mbR^{p_g}$ is the vector of
corresponding regression coefficients of the $g$th group. The
nonconvex group penalized estimator is defined as the minimizer of
\begin{equation}\label{grp:obj}
    Q_{\lambda,\gamma}(\bbeta)=\frac{1}{2n}\|\by-\sum_{g=1}^{G}\bX_g\bbeta_g\|_2^2 + \sum_{g=1}^{G} J_{\lambda_g,\gamma}
    (\|\bbeta_g\|_2),
\end{equation}
where $J_{\lambda_g,\gamma}(\cdot)$ is a penalty function applied to
the $\ell_2$-norm of $\bbeta_g$.  It is common practice
\citep{yuan2006model,Simon2011a} to adjust the regularization
parameter for each group using $\lambda_g=\lambda\sqrt{p_g}$ to
account for differences in group size. By the KKT conditions, the
local minimizer $\hat\bbeta\in\mbR^p$ satisfies
\begin{align}\label{kkt:local}
  \begin{aligned}
    \quad\bX_{g}'&\r/n = \lambda\sqrt{p_g}
    \hat\bbeta_g/\|\hat\bbeta_g\|_2, && g \in \cA,\\
    ~\big|\big|\bX_{g}'&\r/n\big|\big|_2 <
    \lambda\sqrt{p_g}, && g \in \cA^c,\\
  \end{aligned}
\end{align}
where $\cA=\{g:\|\hat\bbeta_g\|_2\neq0\}$ is the index set of
nonzero groups. Similar to standard variable selection, we can
derive strong rules to discard the $g$th group for group MCP and
group SCAD as follows:
\begin{align}
    \text{group MCP :} \left\|\bX_{g}'\r_{k-1}/n\right\|_2
    &< \sqrt{p_g}\left\{\lambda_k + \frac{\gamma}{\gamma-1}(\lambda_k
    - \lambda_{k-1})\right\}\label{rule:gMCP}\\
    \text{group SCAD :} \left\|\bX_{g}'\r_{k-1}/n\right\|_2
    &< \sqrt{p_g}\left\{\lambda_k + \frac{\gamma}{\gamma-2}(\lambda_k -
    \lambda_{k-1})\right\}\label{rule:gSCAD}.
\end{align}
Although we framed this derivation in the linear regression setting,
note that these rules apply to the generalized linear model case as
well, provided that the link function is included in the calculation
of $\r_{k-1}$, as in Section~\ref{glm}.

\section{Simulation study}\label{simulation}

In this section, we carry out a more thorough investigation of the
illustration presented in Figure~\ref{loc-con} in terms of the
frequency of strong rule violations.  We first consider linear and
logistic regression for MCP and SCAD.  The simulation design follows
the description in Section~\ref{strong:non} for the linear regression
case; for logistic regression, the design is the same except that
$y_i$ follows a Bernoulli distribution with the logistic link function
and the nonzero regression coefficients equal $\pm 0.5$.  For 100
independently generated data sets, we record the number of eliminated
variables, the number of $\lambda$ values at which a violation
occurred, and the total number of erroneously discarded variables.
For the violations, we also record whether the violation occurred in
the locally convex region of the solution path or not.

\begin{table}[ht]
  \centering
  \caption{Simulation results for MCP and SCAD strong rules with
    $n=200$ and $p=2,000$.  Results averaged over 100 independent data
    sets.}
    \begin{tabular}{ccccccc}
\hline  \emph{Model} & \emph{Method} & $\rho$
        & \head{2.0cm}{\small Average \# of eliminated variables}
        & \head{1.8cm}{\small Number of violated $\lambda$ values}
        & \head{1.8cm}{\small Number of violated variables}
        & \head{2.3cm}{\small Number of violated $\lambda$ (convex region)}\\
\hline
            &MCP     &0        &1971.17      &1.23      &4.13      &0.01 \\
Linear      &        &0.5      &1973.76      &6.28      &12.74     &0.01 \\
\cline{2-7}
regression  &SCAD    &0        &1958.19      &0.16      &0.62      &-    \\
            &        &0.5      &1958.77      &7.69      &36.37     &-    \\
\hline
            &MCP     &0        &1970.81      &2.36      &6.37      &-    \\
Logistic    &        &0.5      &1982.41      &2.23      &3.98      &-    \\
\cline{2-7}
regression  &SCAD    &0        &1935.91      &3.72      &18.23     &-    \\
            &        &0.5      &1966.88      &4.51      &16.96     &-    \\
\hline
    \end{tabular}\label{tab-var}
\end{table}

Table~\ref{tab-var} presents the results of this simulation, averaged
over the 100 replications.  Overall, the table reflects the earlier
observations made concerning Figure~\ref{loc-con}.  The strong rules
discard a large proportion of variables in the inactive sets and
thereby achieve considerable dimension reduction in $p > n$ problems.
The rules are not foolproof: violations occur regularly, and the
problem is exacerbated by high correlation among the covariates,
although logistic regression is less sensitive to correlation than is
linear regression.

Nevertheless, violations only occur at a small number of the 100
$\lam$ values along the solution path, and almost always occur in the
non-locally convex region.  For example, in the linear regression case
with $\rho=0$, a single violation in 100 data sets was observed for
MCP in the locally convex region, which accounts for less than 1\% of
the observed violations.  In many of the scenarios, no violations in
the locally convex region occurred.

We also study the performance of strong rules for group variable
selection using group MCP and group SCAD using the following
simulation design. The covariates follow a standard Gaussian
distribution with a block-diagonal correlation structure such that
within-block correlation is 0.5. The design matrix consists of 500
groups (blocks), each with 4 elements. The coefficients for the first
6 groups are equal to $\pm 1$ for linear regression and $\pm 0.5$ for
logistic regression; the coefficients in the other 494
groups are all zero.  We fixed the sample size at 200 (i.e., $n=200$,
$G=500$ and $p=2,000$).

\begin{table}[ht]
  \centering
  \caption{Simulation results for group MCP and group SCAD strong
    rules with $n=200,~G=500$ and $p=2,000$.  Within-block correlation
    $\rho$ is 0.5.  Results averaged over 100 independent data sets.}
    \begin{tabular}{cccccc}
\hline  \emph{Model} & \emph{Method}
        & \head{2.0cm}{\small Average \# of eliminated groups}
        & \head{1.8cm}{\small Number of violated $\lambda$ values}
        & \head{1.8cm}{\small Number of violated groups}
        & \head{2.3cm}{\small Number of violated $\lambda$ (convex region)}\\
\hline
Linear      &gMCP    &492.99 &-       &-       &-      \\
regression  &gSCAD   &492.44 &-       &-       &-      \\
\hline
Logistic    &gMCP    &492.07 &0.11    &0.11    &0.01   \\
regression  &gSCAD   &487.30 &1.12    &1.78    &-      \\
\hline
    \end{tabular}\label{tab-grp}
\end{table}

Table \ref{tab-grp} shows the number of discarded groups, as well as
the number of strong rule violations, averaged over 100 independent
data sets.  As in the non-grouped case, violations occur only for a
small fraction of the 100 $\lam$ values along the solution path, and
almost always in the non-locally convex region.

\section{Incorporation of strong rules into model-fitting algorithms}\label{algorithm}

In this section, we discuss the incorporation of strong rules into the
coordinate descent (CD) algorithm of \cite{breheny2011coordinate} for
fitting nonconvex penalized regression models.  The algorithms we
propose may be viewed as modifications of the idea behind coordinate
descent: rather than cycling over the full set of variables with every
iteration, the availability of strong rules and other heuristics allow
one to carry out {\em targeted cycling} in which computational effort
is concentrated on the variables most likely to be nonzero and
therefore change from one $\lam$ value to the next.  We consider three
targeted cycling algorithms: one based on strong rules, one based on
active set cycling, and a hybrid algorithm combining the two
heuristics.

\begin{algorithm}
\caption{\label{Alg:basic} Dimension reduction using strong rules for targeted cycling}
  \begin{algorithmic}
    \For{$k = 1, 2, \ldots, m$}
      \State Calculate the strong set $\cS(\lambda_k)$ and let $\cT=\cS(\lambda_k)$
      \Repeat
        \State Find the solution $\hat\bbeta_{\cT}(\lambda_k)$ using only the variables in $\cT$
        \State Find $\cV=\{j\in\cT^c:|\bx_j'\r/n| \geq \lambda_k\}$
        \State Update $\cT$ by $\cT\cup\cV$
      \Until $\cV=\emptyset$
    \EndFor
  \end{algorithmic}
\end{algorithm}

Algorithm~\ref{Alg:basic} describes the incorporation of strong rules
into the coordinate descent algorithm; we refer to this approach as
the {\em strong rule algorithm}.  The algorithm relies on computing
the {\em strong set} $\cS(\lam)$, which we define as the set of
variables remaining after discarding variables according to the strong
rules (\ref{rule:mcp}, \ref{rule:scad}), and then using this set as
the target set $\cT$ that we cycle over until convergence.  As
discussed previously, it is possible for strong rules to be violated,
and therefore necessary to calculate the set of violations $\cV$ in
order to ensure that all solutions $\hat{\bbeta}$ satisfy the KKT
conditions at convergence.

An alternative approach is to use the active set $\cA(\lam_{k-1})$ as
the target set for calculating the solution at the next step in the
solution path, $\hat{\bbeta}(\lam_k)$.  The algorithm, which we refer
to as {\em active set cycling} \citep{Friedman2010}, is the same as
Algorithm~\ref{Alg:basic} with the active set $\cA(\lam_{k-1})$ replacing the
strong set $\cS(\lam_k)$.

\begin{algorithm}
\caption{\label{Alg:hybrid} Dimension reduction using a hybrid of strong rules and active set cycling}
  \begin{algorithmic}
    \For{$k = 1, 2, \ldots, m$}
      \State Set $\cT=\cA(\lambda_{k-1})$ and $\cS=\cS(\lambda_k)$
      \Repeat
        \Repeat
          \State Find the solution $\hat\bbeta_{\cT}(\lambda_k)$ using only the variables in $\cT$
          \State Find $\cV_1=\{j\in\cS\setminus\cT:|\bx_j'\r/n| \geq \lambda_k\}$
          \State Update $\cT$ by $\cT\cup\cV_1$
        \Until $\cV_1=\emptyset$
        \State Find $\cV_2=\{j\in\cT^c \setminus \cS:|\bx_j'\r/n| \geq \lambda_k\}$
        \State Update $\cT$ by $\cT\cup\cV_2$
      \Until $\cV_2=\emptyset$
    \EndFor
  \end{algorithmic}
\end{algorithm}

The final approach we consider combines the active set and strong sets
into an algorithm that involves two-stage targeted cycling.  The
details are provided in Algorithm~\ref{Alg:hybrid}, which we refer to
as the {\em hybrid algorithm}.

Contrasting the four algorithms (cyclic, strong, active, and hybrid),
there is a tradeoff between how aggressive the algorithms are in terms
of discarding variables and how often violations involving erroneously
discarded variables occur.  Discarding variables naturally increases
the speed of optimization over the target set; however, violations
introduce a computational cost as well, since the iterative targeted
cycling procedure must be restarted and the KKT conditions re-checked.
At one extreme, active set cycling discards the largest number of
variables, but its targeted cycling rule is violated every time a new
variable enters the active set.  On the other extreme, full cyclic
coordinate descent does not have to contend with violations or
re-check any KKT conditions, but must contend with the full set of
variable at every step.  The strong and hybrid algorithms attempt to
occupy a middle ground between these two extremes, reducing
dimensionality as much as possible without introducing a large number
of violations.

\begin{figure}[ht!]
 \centering
 \includegraphics[width=\linewidth]{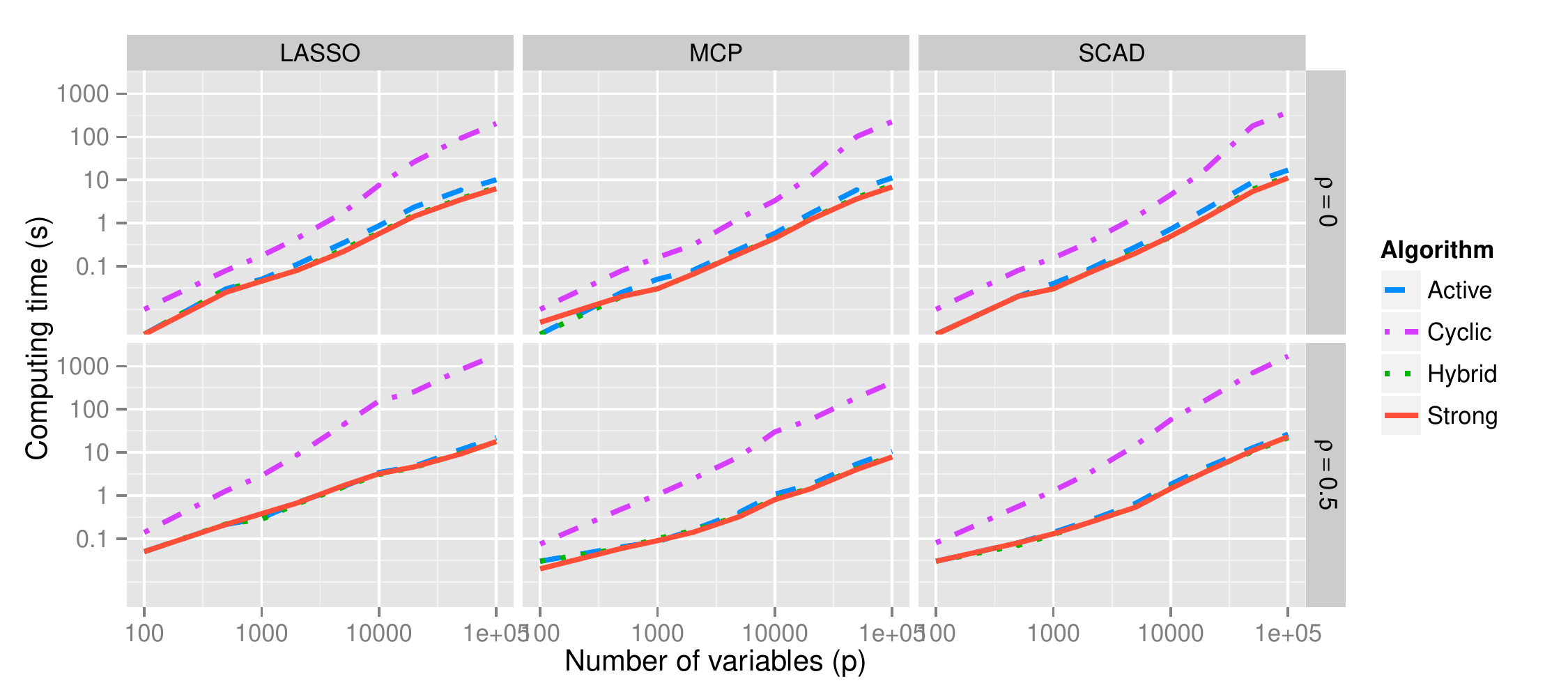}
 \caption{\label{fig-ncv23} \small Comparison of the cyclic CD
   algorithm and targeted cycling algorithms in terms of computing
   time required to fit the entire coefficient path down to
   $\lam_{\min}/\lam_{\max}=0.05$ for linear regression as a function
   of the number of covariates $p$.  Both axes are on the log
   scale. Median times over 20 replications are displayed.}
\end{figure}

Figure~\ref{fig-ncv23} demonstrates that all three targeted cycling
algorithms are considerably faster than full cyclic coordinate
descent, and that the magnitude of the difference is substantial for
high dimensional problems.  For example, the median time required to
fit a SCAD model with $\rho=0.5$ and $p=100,000$ was 1,711 seconds
using cyclic coordinate descent and just 23 seconds using the strong
rule algorithm.  It is worth noting that even though strong rules are
more likely to be violated as correlation increases, the fact that
optimization algorithms must go through a larger number of iterations
in this case results in an even greater advantage for targeted cycling
in the correlated case than in the uncorrelated case.


\begin{figure}[ht!]
 \centering
 \includegraphics[width=\linewidth]{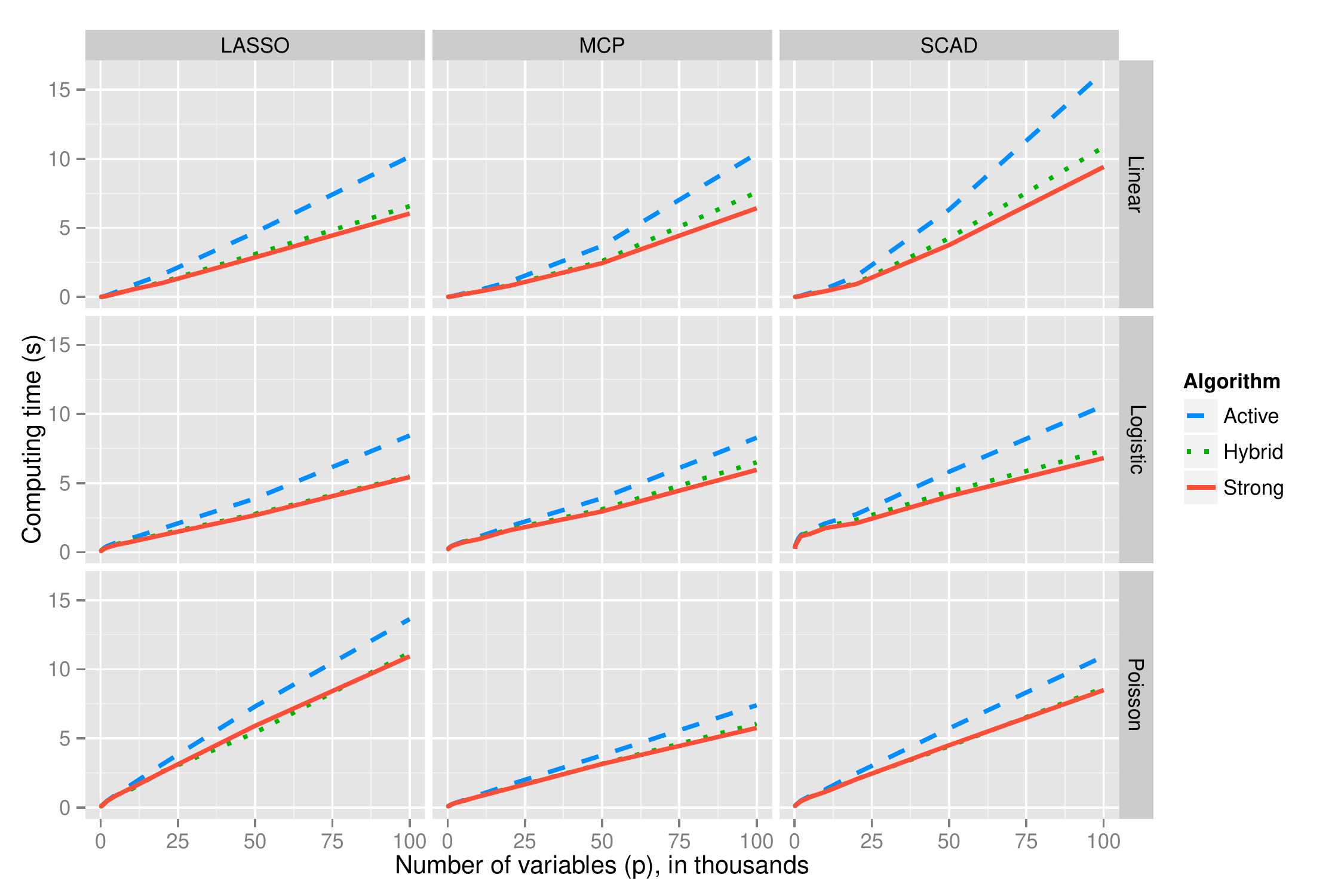}
 \caption{\label{fig-time} \small Comparison of targeted cycling
   algorithms in terms of computational time required to fit the
   entire coefficient path for linear (top), logistic (middle) and
   Poisson regression (bottom).  Median computing times over 20
   replications are displayed.}
\end{figure}

In Figure~\ref{fig-ncv23}, it is clear that targeted cycling is more
efficient than full cyclic CD, but it is unclear how the target
cycling algorithms compare to each other.  In Figure~\ref{fig-time},
we compare the speed of the three targeted cycling algorithms for
linear, logistic, and Poisson regression.  In each case, 20 variables
are set to $\pm 0.5$ with the remaining variables set to zero, and the
outcome follows the distribution assumed in the GLM.  In all cases,
the strong rule and hybrid algorithms were seen to be more efficient
than active set cycling.  Although the difference in computing time
between the targeted cycling algorithms is minor for small $p$, active
set cycling can take almost twice as long for high-dimensional models.
For example, fitting a SCAD-penalized logistic regression model with
$p=100,000$ required a median computing time of 11 seconds for active
cycling and only 7 seconds for the strong rule and hybrid algorithms.

\begin{figure}[ht!]
 \centering
 \includegraphics[width=\linewidth]{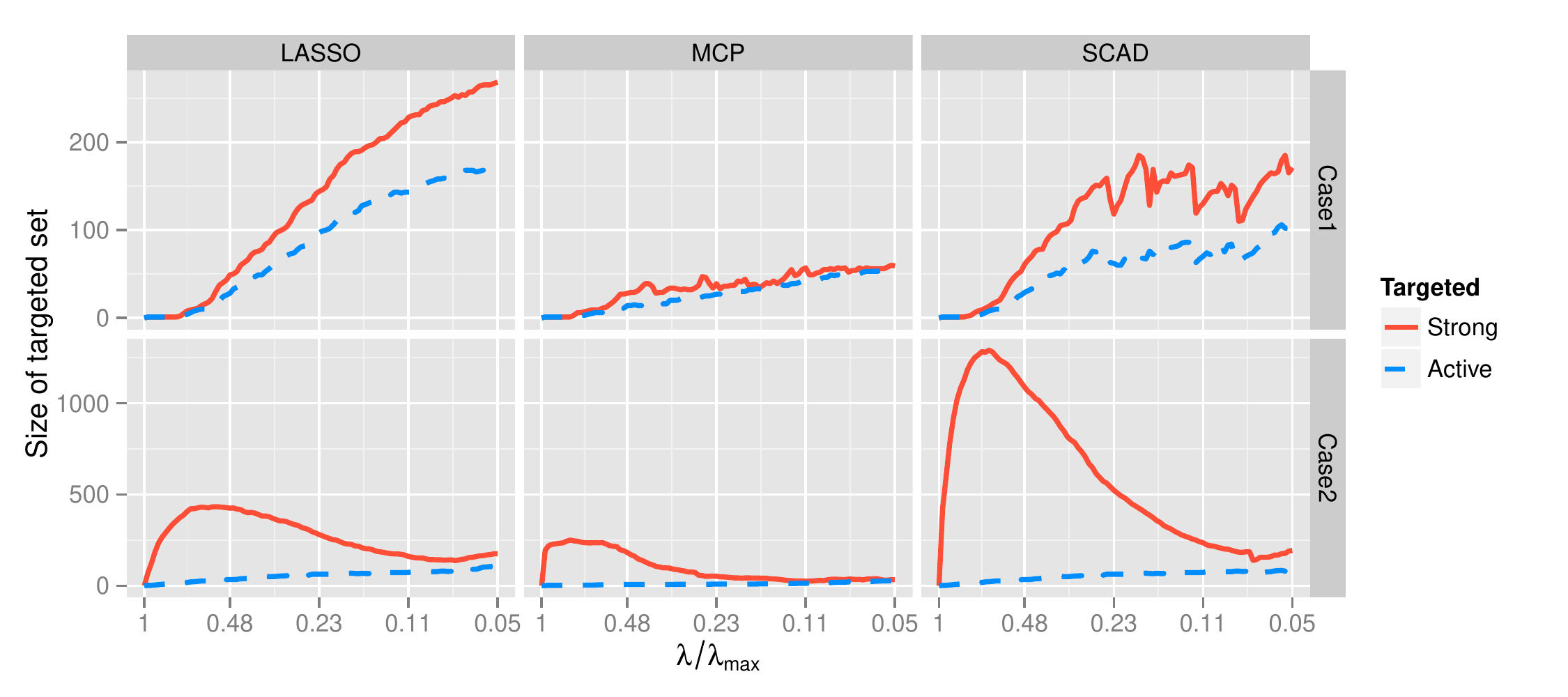}
 \caption{\label{str-act} \small Comparison of the size of strong and
   active sets for each $\lambda$ for simulated Gaussian data with
   $\rho=0$, $n=200$ and $p=20,000$.  In ``Case 1'' (top panel),
   nonzero coefficients are equal to $\pm 1$; in ``Case 2'' (bottom
   panel), all nonzero coefficients equal $+1$.}
\end{figure}

Although the strong rule algorithm was slightly faster than the hybrid
algorithm in Figure~\ref{fig-time}, we have found that there are
situations in which the hybrid algorithm offers considerably better
performance than the strong rule approach.  We depict one such
situation for linear regression in Figure~\ref{str-act}.  The top
panel (``Case 1'') of Figure~\ref{str-act} is similar to the
situations we have examined so far, with $n=200$, $p=20,000$,
$\rho=0$, and 20 nonzero coefficients equal to $\pm 1$.  Here, the
variance of Gaussian noise was chosen so that the signal-to-noise
ratio was equal to 3.  The setting for the bottom panel (``Case
2'') is the same, except that the nonzero coefficients all have
coefficients equal to $+1$.  In Case 1, the size of the target set for
the strong rule algorithm matches the active set quite closely, and
nearly all the variables that can be eliminated are eliminated by the
strong rules.  In Case 2, however, although the strong rules are still
valid and rarely violated, they do not yield a target set that closely
matches the active set, and fail to discard hundreds of variables that
remained inactive.

In Case 1, there were minimal differences between the computing time
of the three targeted cycling algorithms (all were within 1 second of
each other).  In Case 2, however, the strong rule algorithm was
substantially slower than active cycling and the hybrid algorithm due
to the much larger size of its target set.  For SCAD, where the
difference in target set size was most dramatic, the strong rule
algorithm required 19 seconds, while active cycling required just 5.
As it is designed to do, the hybrid algorithm utilizes the best
features of each heuristic and requires just 4 seconds to compute the
solution path.

In summary, we find the hybrid algorithm to be the most robust of the
targeted cycling approaches -- never much slower than the strong rule
algorithm, and in some cases much faster.  For this reason, we have
implemented the hybrid algorithm for lasso, SCAD, and MCP-penalized
linear, logistic, and Poisson regression in the {\tt ncvreg} package.

\section{Application to genome-wide association studies}
\label{Sec:gwas}

In this section, we apply the algorithms described in
Section~\ref{algorithm} to real data from a genome-wide association
study (GWAS) of preeclampsia.  The data were collected during the
Study of Pregnancy Hypertension in Iowa (SOPHIA), a population-based
case-control study.  We provide a brief description of the data
here; the study is described in greater detail in \citet{Zhao2012}.

The sample consists of 177 mothers diagnosed with preeclampsia
according to National Heart, Lung and Blood Institute guidelines and
115 mothers with normal blood pressure to serve as controls.  All
292 mothers were genotyped using the Affymetrix Genome-Wide Human
SNP Array 6.0 (Affymetrix, Santa Clara, CA).  After applying quality
control procedures and eliminating monomorphic markers, we were left
with 810,198 single-nucleotide polymorphisms (SNPs) to serve as
potential predictors of preeclampsia risk.

We analyzed this data using MCP-penalized logistic regression with
case-control status as the response variable.  Allele effects were
assumed to be additive and independent, thereby yielding a design
matrix with $n=292$ and $p=810,198$.  Due to the fact that $p \gg
n$, we fit the penalized regression model over a relatively small
portion of the coefficient path, down to
$\lambda_{\min}/\lambda_{\max}=0.8$, at which point 22 SNPs had
entered the model.  Despite the large number of features in the
design matrix, the penalized regression models could be fit very
rapidly: using the strong rule algorithm, the solution path could be
fit in just 4.7 seconds on a standard desktop computer (3.60GHz
Intel Xeon processor, 16 GB RAM).  The active cycling algorithm took
somewhat longer, at 7.9 seconds, while the performance of the hybrid
algorithm was similar to that of the strong rule approach (4.9
seconds to fit the solution path).

The SNPs selected by the penalized regression model are consistent
with the top-ranked SNPs in terms of univariate hypothesis testing
using Fisher's exact test, as reported in \citet{Zhao2012}. However,
we reach the same conclusion that the authors of the previous study
reached -- namely, that there is insufficient evidence in the data
to perform variable selection with any meaningful degree of
reliability.  In particular, when we carry out 10-fold
cross-validation for the purposes of selecting $\lambda$, we find
that the optimal model is the intercept-only model.

Although this particular study was negative in terms of identifying
genetic risk factors for preeclampsia, it illustrates the
feasibility of fitting penalized regression models to very
high-dimensional data. The current genome-wide association
literature is overwhelmingly focused on univariate tests, which have
many shortcomings compared to multivariate modeling: inefficiency,
increased risk of confounding, and limited predictive inference,
among others.  Several authors have recommended penalized regression
as an alternative, and discussed its benefits in comparison with
univariate testing \citep{Zhou2010a,Wu2009}.  Others, however, have
judged the problem to be computationally impractical for the very
high dimensions that prevail at the genome-wide scale and developed
multi-stage or iterative screening proposals to reduce the
dimensionality of the problem \citep{Fan2008,Shi2011,Zhao2012a}.  We
demonstrate here that such approaches are not necessary -- or,
depending on your perspective, that screening is indeed a very
useful idea, but it can be incorporated directly into coordinate
descent algorithms through targeted cycling.

\section{Discussion}

Concern over the computational burden of penalized regression in
very high dimensions has prevented its use in many fields,
particularly in genetics.  This concern, in turn, has led many
researchers to pre-screening procedures to reduce the dimensionality
of the problem before fitting the penalized regression model.  At
best, this complicates both the theoretical study of such procedures
and the practical implementation of procedures such as
cross-validation.  At worst, it opens the door for bad statistical
practice by obfuscating the multiple comparison problem.  For
example, if pre-screening is used to select candidate variables on
the full data set, and then cross-validation is used to select a
tuning parameter $\lambda$, the resulting inference is heavily
biased by the fact that the external validation data is not truly
external, as it has already been used for screening.

It is possible to carry out unbiased cross-validation in the presence
of screening, but it is also very easy for a well-intentioned
investigator to make a mistake \citep[a thorough discussion of this
  issue may be found in][]{Hastie2009}. In contrast, cross-validation
is both straightforward and computationally feasible, and already
implemented existing software such as {\tt glmnet} and {\tt ncvreg}.
In particular, for the analysis in Section~\ref{Sec:gwas}, ten-fold
cross validation was carried out in under a minute despite fitting
nonconvex penalized logistic regression models with $p=810,198$
variables.

With this work, we have demonstrated that fitting high-dimensional
nonconvex penalized regression models can be made computationally
feasible through the use of targeted cycling and strong rules to
achieve dimension reduction.  Furthermore, by sharing implementations
of these algorithms in the publicly available {\tt R} package {\tt
  ncvreg}, we hope to encourage researchers to adopt these methods
with greater regularity for analyzing high-dimensional data.

\bibliographystyle{ims}

\end{document}